\documentclass[11pt]{article}
\pdfoutput=1

\usepackage{euscript}
\usepackage{amssymb}
\usepackage{amsfonts}
\usepackage{amsbsy}
\usepackage{epsfig}
\usepackage{amsthm}
\usepackage{amscd}
\usepackage{amstext}
\usepackage{verbatim}
\usepackage{amsmath}
\usepackage{cancel}
\usepackage{capt-of}

\def\ben{\begin{equation}}
\def\een{\end{equation}}

\let\a=\alpha   \let\d=\delta 
   \let\k=\kappa
\let\l=\lambda     
\let\s=\sigma \let\t=\tau

\let\w=\omega \let\G=\Gamma

\let\pa=\partial

\newcommand{\ba}{\begin{align}}
\newcommand{\ea}{\end{align}}

\newcommand{\bi}{\begin{itemize}}
\newcommand{\ei}{\end{itemize}}
\newcommand{\specialcell}[2][c]{%
  \begin{tabular}[#1]{@{}c@{}}#2\end{tabular}}
\newcommand{\rchi}{{\mathpalette\irchi\relax}}
\newcommand{\irchi}[2]{\raisebox{\depth}{$#1\chi$}}

\def\be{\begin{equation}}
\def\ee{\end{equation}}
\def\beq{\begin{equation}}
\def\eeq{\end{equation}}

\def\dalemb#1#2{{\vbox{\hrule height .#2pt
       \hbox{\vrule width.#2pt height#1pt \kern#1pt
               \vrule width.#2pt}
       \hrule height.#2pt}}}

\newcommand{\bea}{\begin{eqnarray}}
\newcommand{\eea}{\end{eqnarray}}

\def\vep{{\varepsilon}}

\def\ocal{{\mathcal{O}}}

\def\cQ{\mathcal{Q}}

\begin{document}

\begin{center}

{ \Large {\bf Non-Fermi liquids and the Wiedemann-Franz law
}}

\vspace{1cm}

Raghu Mahajan, Maissam Barkeshli, Sean A. Hartnoll

\vspace{1cm}

{\small
{\it Department of Physics, Stanford University, \\
Stanford, CA 94305-4060, USA }}

\vspace{1.6cm}

\end{center}

\begin{abstract}

A general discussion of the ratio of thermal and electrical conductivities in non-Fermi liquid metals is given.
In metals with sharp Drude peaks, the relevant physics is correctly organized around the slow relaxation of almost-conserved momenta. While in Fermi liquids both currents and momenta relax slowly, due to the weakness of interactions among low energy excitations, in strongly interacting non-Fermi liquids typically only momenta relax slowly. It follows that the conductivities of such non-Fermi liquids are obtained within a fundamentally different kinematics to Fermi liquids. Among these strongly interacting non-Fermi liquids we distinguish cases with only one almost-conserved momentum, which we term quasi-hydrodynamic metals, and with many patchwise almost-conserved momenta. For all these cases, we obtain universal expressions for the ratio of conductivities that violate the Wiedemann-Franz law. We further discuss the case in which long-lived `cold' quasiparticles, in general with unconventional scattering rates, coexist with strongly interacting hot spots, lines or bands. For these cases, we characterize circumstances under which non-Fermi liquid transport, in particular a linear in temperature resistivity, is and is not compatible with the Wiedemann-Franz law. We suggest the likely outcome of future transport experiments on CeCoIn$_5$, YbRh$_2$Si$_2$ and Sr$_3$Ru$_2$O$_7$ at their critical magnetic fields.

\end{abstract}

\pagebreak
\setcounter{page}{1}

\section{The Wiedemann-Franz law}

The Wiedemann-Franz law for the Lorenz ratio of thermal conductivity $\k$ to electrical conductivity
$\sigma$, at low temperature $T$,
\be\label{eq:wf}
L \equiv \frac{\k}{\s T} = \frac{\pi^2}{3} \equiv L_0 \,,
\ee
in units with $k_B = e = 1$, is a robust feature of Fermi liquids at low temperature \cite{ziman}. In a non-Fermi liquid metal, there is ample reason to expect the Wiedemann-Franz law to be violated, see e.g. \cite{pepin}. There may be additional gapless neutral collective degrees of freedom present that transport heat but not charge. There may furthermore be inelastic scattering between charged and neutral degrees of freedom that affects heat and charge transport differently. It may therefore seem surprising that in several systems exhibiting supposedly hallmark non-Fermi liquid phenomenology, such as a linear in temperature resistivity, the Wiedemann-Franz law is observed to hold at the lowest temperatures \cite{wf1,wf2,wf3,wf4,wf5}, while the interpretation of recently reported violations of the law in such materials has proved contentious \cite{con1,con1a, con2}. On the other hand, violations of the Wiedemann-Franz law in metallic regimes at low temperatures have been reported in the underdoped cuprates \cite{nwf1,nwf2,nwf3}, upon suppressing superconductivity, and also in the c-axis conductivity of a heavy fermion non-Fermi liquid \cite{caxis}. This panoply of results clearly presents a theoretical challenge.

The thermal and electrical conductivities are subtle observables because they depend crucially on momentum relaxation in order to be finite. The interactions between charge carriers and gapless neutral degrees of freedom causing non-Fermi liquid behavior may not in themselves cause momentum relaxation. To understand the ratio of conductivities, one must therefore characterize the interplay between momentum-conserving and momentum-non-conserving interactions. If a quasiparticle description is apposite, this question can be addressed via the Boltzmann equation, e.g. \cite{ziman, maslov}. In this paper we shall make some simple observations about the ratio of thermal to electric conductivity in generic non-Fermi liquids. Important differences will be found between cases in which weakly interacting quasiparticles are present and those in which they are not. The tool we shall use to organize our discussion, which emphasizes the role of momentum relaxation, is the memory matrix formalism \cite{forster}.

The memory matrix formalism allows a clear-headed discussion of conductivities without recourse to a quasiparticle worldview. As such it has proved useful in studying the transport of one dimensional interacting electrons \cite{ar1,ar2} and higher dimensional systems at quantum criticality \cite{Hartnoll:2007ih, Hartnoll:2008hs, Hartnoll:2012rj}. In particular, a memory matrix approach successfully predicted large violation of the Wiedemann-Franz law in one dimensional Luttinger liquids \cite{oned, oned2}.
The insight exploited by the memory matrix approach is the following: Even if the system is strongly interacting, in a metallic phase the effects of momentum relaxation (due to quenched impurities or a lattice) on low energy processes can often be understood as a perturbation about an effective translation-invariant low energy theory. In many cases this will be synonymous with the system exhibiting a well-defined Drude peak in the optical conductivity at the low temperatures of interest. A Drude peak certainly does not imply a quasiparticle description of transport. The memory matrix method then proceeds to isolate precisely the quantities that may be treated perturbatively in these cases.

The memory {\it matrix} formalism is to be distinguished from the earlier `memory {\it function} approximation' developed in e.g. \cite{gw,pl,gia}. In that approach one works perturbatively in the relaxation of the electrical current rather than the momentum. This is most appropriate for non-interacting systems in which the total current $\vec J$ itself is a conserved operator (i.e. $\dot {\vec J} = 0$) prior to impurity and other scatterings. For the strongly interacting non-Fermi liquids we wish to consider below, the total current itself is not an almost conserved operator and one should instead focus on momentum conservation. The momentum and current operators are typically not equal in these systems. The memory matrix furthermore includes various projection operators, see for instance equation (\ref{eq:memory}) below, that render it an exact expression rather than an approximation.

In this paper we focus on a rough classification of the different possible kinematic regimes that can control transport in the theories of interest. Our results take the form of relationships between different physical quantities that pertain depending on the kinematics of the system. For instance,
part of the power of the Wiedemann-Franz law (\ref{eq:wf}) is that the right hand side is a pure number that is computable using Fermi liquid theory. Effects sensitive to processes external to Fermi liquid theory, in particular the quasiparticle lifetime, have cancelled out of the final expression. We shall give a memory matrix description of this fact below, and, from our more general perspective, establish conditions under which an analogous universality may be achieved for non-Fermi liquids.

Our considerations apply to the following cases:

\begin{enumerate}

\item {\it Non-Fermi liquids with no quasiparticles}. Section \ref{sec:nfl}. These may include two dimensional nematic quantum critical points and Fermi surfaces coupled to emergent gauge fields, in which the full Fermi surface goes critical. The kinematics of almost-conserved quantities is entirely different to Fermi liquid theory and the Wiedemann-Franz law is expected to not hold even approximately. Currents are not approximately conserved. The only almost-conserved quantities are momenta, of which there may be one or many (if the patches of a strongly interacting Fermi surface decouple). For the case of only one conserved momentum, we refer to these as `quasi-hydrodynamic metals', we obtain
the ratios of conductivities
\be\label{eq:newWF}
\frac{\overline \k}{\s T} = \frac{1}{T^2} \frac{\rchi_{QP}^2}{\rchi_{JP}^2} \qquad \text{and} \qquad \frac{\k}{\s T} \ll 1\,.
\ee
Here $\overline \kappa$ is the thermal conductivity at zero electric field, while $\k$ is the conductivity at zero electric current. The $\rchi_{AB}$ are static susceptibilities involving the operators for the total momentum $\vec P$, electric current $\vec J$ and heat current $\vec Q$. While the specific values  and temperature dependence of these susceptibilities will depend on details of the theory at all energy scales, the left hand result in (\ref{eq:newWF}) is universal in the sense that the mechanism of momentum relaxation has once again cancelled from the ratio of conductivities. This universality is analogous to the Wiedemann-Franz law. Cases with many patchwise conserved momenta are discussed in section \ref{sec:nflmany}, where we obtain
\be
\frac{{\overline \kappa}}{\sigma T} \sim \frac{\kappa}{\sigma T} \sim \frac{1}{T^2} \left\langle \frac{\rchi_{PQ}^2}{\rchi_{PJ}^2} \right\rangle
 \,.
\ee
While $\k = {\overline \k}$ in a Fermi liquid, $\k \sim {\overline \k}$ showing the same temperature scaling without
equality is a diagnostic for strongly interacting transport with patchwise conserved momenta.
The angled brackets refer to a specific average over the Fermi surface.

\item {\it Non-Fermi liquids with long-lived quasiparticles}. Section \ref{sec:long}. These are systems in which some or all of the charge and heat carriers remain long lived, despite the system exhibiting unconventional metallic transport.
These may include finite wavevector quantum critical points, such as spin and charge density wave transitions in metals, which have hot and cold fermionic excitations. They may also include situations where one electron band goes critical but others do not. Finally these include cases in which the whole Fermi surface is critical but the fermions retain a quasiparticle character. The Wiedemann-Franz law will be obeyed in these systems at low temperatures if the long-lived electronic quasiparticles dominate the charge and heat transport at low temperatures and scatter elastically. If the electronic quasiparticles have an unconventional scattering rate, the Wiedemann-Franz law will coexist with non-Fermi liquid transport. We highlight a scenario in which the Wiedemann-Franz law can coexist at low temperatures with a  linear in temperature electrical resistivity: if the `cold' degrees of freedom scatter elastically off the `hot' modes, which in turn relax momentum more efficiently than the cold excitations. In these circumstances, the hot modes act as `generalized phonons'.

\end{enumerate}

The above outline makes clear that the status of the Wiedemann-Franz law at low temperatures in a non-Fermi liquid provides immediate insight into the nature of the excitations in the system. In section \ref{sec:data} we use the simple observations summarized above to organize the existing experimental data on the Wiedemann-Franz law in heavy fermions, ruthenates and the cuprates. See table \ref{tab:mat} below. The law has not yet been studied in certain natural temperature regimes, or ambiguous results have been obtained, allowing us to predict the outcome of future measurements.

\section{The memory matrix}

Let us briefly introduce the memory matrix. To follow the present paper, only the logical structure of equation (\ref{eq:sw}) below need be understood. We will avoid actual evaluation of the memory matrix. The starting point are the correlators
$\rchi(z)$ and the static susceptibility $\rchi$
\be\label{eq:chi}
\rchi_{AB}(z) = i \int_0^\infty dt e^{i z t} \langle [A(t), B(0)] \rangle \,, \qquad \rchi_{AB} = \lim_{z \to i0^+} \rchi_{AB}(z) \,,
\ee
with the operators $A,B$ being taken from the set of (i) the operators whose two point function we wish to compute, these are the total electric and thermal currents $\{ \vec J, \vec Q \}$, and (ii) any almost-conserved operators that have an overlap with the currents of interest, for instance the total momentum $\vec P$. The quantity of interest to us is the matrix of conductivities
\be
\sigma(\w) = \frac{\rchi(\w) - \rchi}{i \w} \,.
\ee
The limit of real frequencies is $z \to \w + i0^+$, as this is where the integrals in (\ref{eq:chi}) converge.
The memory matrix formalism expresses the matrix of conductivities as
\be\label{eq:sw}
\sigma(\w) = \frac{1}{- i \w + M(\w) \rchi^{-1}} \rchi \,,
\ee
with the memory matrix being given by \cite{forster}
\be\label{eq:memory}
M_{AB}(\w) = \int_0^{1/T} d\lambda \left\langle \dot A(0) \cQ \frac{i}{\w - \cQ L \cQ} \cQ \dot B(i\l) \right\rangle \,.
\ee
Here $L$ is the Liouville operator $L = [H,\cdot]$ and $\cQ$ projects onto the space of operators orthogonal to $\{\vec P, \vec J, \vec Q\}$. The reason it is useful to introduce the memory matrix is that if the effects of momentum or current non-conservation are small at low energies, this quantity can be treated perturbatively because the $\dot A$ and/or $\dot B$ appearing in (\ref{eq:memory}) will be small. The d.c. conductivities are
\be\label{eq:gam}
\sigma^\text{d.c.} = \lim_{\w \to 0} \rchi M(\w)^{-1} \rchi \equiv \Gamma^{-1} \rchi \,.
\ee
Here we introduced the matrix of relaxation rates $\Gamma$. While $\rchi$ and $M$ will be symmetric, $\Gamma$ need not be. Equation (\ref{eq:gam}) expresses the d.c. conductivities as a combination of `fast' processes described by $\rchi$, in which currents and momenta source each other, and `slow' processes described by $M$. The inverse memory matrix $M^{-1}$ will be dominated by the decay of any almost-conserved quantities in the system that overlap with the currents.
 The quantities of interest are then the electrical, heat and thermoelectric conductivities
\be
\sigma \equiv \sigma^\text{d.c.}_{JJ} \,, \qquad \overline \kappa \equiv \frac{\sigma^\text{d.c.}_{QQ}}{T} \,, \qquad \alpha \equiv \frac{\sigma^\text{d.c.}_{JQ}}{T}  \,.
\ee

Equation (\ref{eq:sw}) shows that when some of the relaxation rates can be treated pertubatively -- when the memory matrix approach is useful -- then there will be a sharp Drude peak in the frequency dependent conductivity.

\section{Fermi liquids}
\label{sec:FL}

As a first application of the memory matrix formalism, we can give a re-derivation of the Wiedemann-Franz law (\ref{eq:wf}). The effective low energy and momentum-conserving theory in this case is the patchwise description of the excitations of a Fermi surface \cite{pol,shank}. Let us label the patches by $\theta$ and the momentum locally perpendicular to the Fermi surface as $k$. Fermi liquids have the rather special property that to lowest order in energies the different patch theories are decoupled and furthermore free. This last statement requires qualification in two dimensions due to the logarithmic growth in the strength of scattering by disorder and by soft collective particle-hole excitations at low temperatures. We return to this point at the end of the section. Decoupled free patches leads to the fact that the quasiparticle densities $\delta n_{\theta k} = c^\dagger{}_{\theta k}  c_{\theta k}$ are all independently conserved. The electron-electron interactions that can relax the density within patches, such as normal forward scattering and umklapp processes, are negligible at the lowest temperatures and energy scales \cite{ziman}. The leading density (and, shortly, momentum) relaxing interaction is elastic scattering by quenched impurities. Thus we write
\be\label{eq:density}
\frac{d}{dt} \delta n_{\theta k} = - \int d\theta' d k' \, \Gamma_{\theta \theta'  k k'} \delta n_{\theta' k'} \,.
\ee
An explicit expression can be obtained for the relaxation rates per unit perpendicular momentum $\Gamma_{\theta \theta' k k'}$ from e.g. the appropriate Boltzmann equation \cite{ziman}. To implement the memory matrix method, we need to assume that the scattering rate $\G$ is small in units of the Fermi energy $\mu$. The Wiedemann-Franz law can in fact hold without this assumption \cite{fullWF}. The large number of almost conserved quantities $\delta n_{\theta k}$ means that the memory matrix method is applied in a particular way for Fermi liquids. Our derivation of Wiedemann-Franz will parallel very closely the usual derivation \cite{ziman}. We are emphasizing the role of various almost conserved quantities from the perspective of the effective low energy theory, with a view to generalizing to the non-Fermi liquid case in the following.

The momentum and heat and electrical currents may be constructed patchwise out of the almost conserved densities. To lowest order in energies
\be\label{eq:PJQ}
\vec P_\theta =  \vec k_{F \,\theta} \int dk \, \delta n_{\theta k} \,, \quad \vec Q_\theta = \vec v_{F \, \theta} \int dk \, \vep_{\theta  k}  \, \delta n_{\theta k} \,, \quad
\vec J_\theta = \vec v_{F \, \theta} \int dk \, \delta n_{\theta k} \,.
\ee
The quasiparticle energy $\vep_{\theta k}$ vanishes at the Fermi surface and so must be kept inside the $k$ integral. The total momenta and currents are obtained by integrating these expressions over the whole Fermi surface.

From (\ref{eq:PJQ}) we that, for this Fermi liquid case, the rate at which the total almost-conserved quantities relax will be controlled by the relaxation of the quasiparticles densities (\ref{eq:density}). We must therefore consider the memory matrix for these densities directly. We can then write, picking some spatial direction $\vec n$ and setting $v_{F \,\theta}  = \vec v_{F \,\theta} \cdot \vec n$,
\bea
\overline \kappa & = & \frac{1}{T} \int d\theta d\theta' dk dk' \, v_{F \,\theta} \, v_{F \,\theta'} \, \vep_{\theta k} \,  \vep_{\theta' k'} \,\sigma^\text{d.c.}_{\delta n_{\theta k}  \delta n_{\theta' k'} } \nonumber \\
& = & \frac{1}{T} \int d\theta d\theta'  v_{F \,\theta} \, v_{F \,\theta'} \int dk dk' \left(\G^{-1}\right)_{\theta \theta' k k'} 
\vep_{\theta k} \, \vep_{\theta' k'} \,
 \rchi_{\delta n_{\theta' k'}  \delta n_{\theta' k'}  } \label{eq:kappaline2} \,.
\eea
In the first line $\sigma^\text{d.c.}_{\delta n \delta n}$ is defined by the general formula (\ref{eq:gam}) for d.c. `conductivities'. In the second line we have used the fact that to lowest order in energies the quasiparticle susceptibility is only nonzero for fluctuations on the same patch and with the same momentum. At the end of this section we mention the role of scattering by collective particle-hole excitations that can couple patches. We have also introduced the matrix of relaxation rates $\G$, as defined in (\ref{eq:gam}), which is the same as the matrix appearing in (\ref{eq:density}). These matrices are seen to be the same upon Fourier transforming the conductivity matrix (\ref{eq:sw}) and taking the late time limit \cite{forster}. Elasticity of the collisions, together with the fact that $\G$ becomes independent of the perpendicular momentum to lowest order in energies, allows us to write $\left(\G^{-1}\right)_{\theta \theta' k k'} = \left(\G^{-1}\right)_{\theta \theta'} \delta(\vep_{\theta k} - \vep_{\theta' k'})$. We may therefore use the energy conservation delta function to replace the $\vep_{\theta k}$ in (\ref{eq:kappaline2}) with $\vep_{\theta' k'}$ to obtain
\bea
\overline \kappa & = &\frac{1}{T} \int d\theta d\theta'  v_{F \,\theta} \, v_{F \,\theta'} \left(\G^{-1}\right)_{\theta \theta'} \int  dk' \vep_{\theta' k'}^2
 \rchi_{\delta n_{\theta' k'}  \delta n_{\theta' k'} }  \int dk \, \delta(\vep_{\theta k} - \vep_{\theta' k'}) \nonumber \\
& = & \frac{T \pi^2}{3} \int d\theta d\theta'  v_{F \,\theta} \, v_{F \,\theta'} \left(\G^{-1}\right)_{\theta \theta'} \int dk'  \rchi_{\delta n_{\theta' k'}  \delta n_{\theta' k'} } \int dk \, \delta(\vep_{\theta k} - \vep_{\theta' k'}) \nonumber \\
& = & \frac{T \pi^2}{3} \sigma \,. \label{eq:derivewf} 
\eea
To obtain the second line we have used the well-known result relating the heat and electric current susceptibilities for free fermions. Namely $\int dk \vep^2 \rchi_{\delta n \delta n} = T^2 \pi^2/3 \int dk \rchi_{\delta n \delta n}$. This follows from performing integrals of the Fermi-Dirac distribution, because the susceptibilities $\rchi_{\delta n \delta n} = f'_\text{FD}(\vep)$, and holds patchwise in a Fermi liquid.

The thermal conductivity at zero electric current is $\kappa = \overline \kappa - \alpha^2 T/\sigma$, with $\a$ the thermoelectric conductivity. The thermoelectric conductivity is given by adapting the formulae in (\ref{eq:kappaline2}) in the obvious way. The susceptibility integral that appears is now $\int dk \vep \rchi_{\delta n \delta n}$ which, again using the Fermi-Dirac distribution, is seen to go like $T^2$ at low temperatures. It follows that $\overline \kappa \gg \alpha^2 T/\sigma$ at low temperatures, and hence $\kappa = \overline \kappa$ in this regime. The result (\ref{eq:derivewf}) therefore implies the Wiedemann-Franz law (\ref{eq:wf}).

In two space dimensions, Fermi liquid theory suffers from marginally relevant perturbations due to disorder and scattering by collective particle-hole excitations. The collective modes of an interacting disordered Fermi liquid lead, in two dimensions, to logarithmic corrections to the electric and thermal conductivities that grow as the temperature is lowered. These corrections violate the Wiedemann-Franz law, e.g. \cite{log1, log2}. In the good metallic regimes we are focussing on, away from metal-insulator transitions, these effects only become important at exponentially small temperatures.

\section{Non-Fermi liquids without quasiparticles}
\label{sec:nfl}

In a Fermi liquid, the small effects of interactions at low energies implied the existence of an infinite collection of conserved densities $\delta n_{\theta k}$ in the effective low energy theory. The non-Fermi liquids that we consider in the remainder are characterized by the persistence of strong interactions in the low energy theory. Therefore, the large number of conserved densities $\delta n_{\theta k}$ are not present. It follows that generically the total electrical and heat currents, $\vec J$ and $\vec Q$, are not conserved (unlike in a Fermi liquid).

We will require an essential additional feature of the non-Fermi liquids. Namely, that in the strongly interacting effective low energy theory, the total momentum is conserved, so that $\dot{\vec P} = 0$ up to the effect of irrelevant or weak marginal operators. This will allow these systems to be `good' metals with a sharp Drude peak, despite the absence of quasiparticles. The kinematics of conserved quantities underlying such non-Fermi liquid transport is significantly different to that of a Fermi liquid. In these non-quasiparticle based circumstances, overly na\"ive analogies such as considering scattering by quantum critical bosons to be similar to scattering by phonons in a Fermi liquid will lead to incorrect conclusions.

The conservation of momentum up to effects that are small at low energies is a key assumption that allows the memory matrix method to work. It can be expected to be true in a metal exhibiting a well-defined Drude peak at low temperatures.
This is because a perturbative approach to the memory matrix will lead to a sharp Drude peak in (\ref{eq:sw}).
Our discussion in this paper does not apply to cases in which momentum-non-conserving interactions have strong effects at low energies. For instance, experimentally speaking, at the boundary of metal-insulator transitions one can encounter `bad' metals that violate the Mott-Ioffe-Regal resistivity bound \cite{Emery:1995zz} and do not exhibit Drude peaks \cite{hussey}. In appendix \ref{sec:strong} we briefly discuss theoretical circumstances in which such strong momentum-violating interactions arise.

While interactions in a non-Fermi liquid mean that the low energy theory is not free, it is still possible that there may be a decoupling of excitations into patches in momentum space, analogous to the patches of a Fermi surface in Fermi liquid theory. In these cases there will not just be one conserved momentum $\vec P$, but rather a family of conserved momenta $\vec P_\theta$, labelled by the patch $\theta$. We will consider these two cases, with one and with many conserved momenta, separately.

\subsection{Quasi-hydrodynamic non-Fermi liquid metals}

We will refer to metallic systems in which there is only one almost-conserved vector operator in the effective low energy theory, typically the total momentum, as `quasi-hydrodynamic' non-Fermi liquids. For such non-Fermi liquids we can obtain general results concerning the ratio of conductivities. The total momentum is relaxed on a much longer timescale than all other quantities, including the currents; schematically $\langle \dot{\vec P} \, \rangle \sim \epsilon \, \mu \, \langle \vec P \, \rangle$, while, for instance, $\langle \dot{\vec J} \, \rangle \sim \mu \langle \vec J \, \rangle$, with $\epsilon \ll 1$ and $\mu$ a microscopic scale. Our discussion is easily adapted to cases in which the electric current is equal to the momentum. The only general requirement for our results is that there be only one almost-conserved vector operator. From the hierarchy of relaxation rates, we can anticipate that in the $\w = 0$ memory matrix $M_{PP} \sim \epsilon^2$ and $M_{PJ} \sim M_{PQ} \sim \epsilon$, while the remaining components of the memory matrix (\ref{eq:memory}) are order one. It follows that the inverse of the memory matrix will be dominated by $\left(M^{-1}\right)_{PP} \sim \epsilon^{-2} \gg 1$.

An explicit example of the above logic was discussed in \cite{Hartnoll:2012rj}. Lattice scattering is included in the low energy theory via coupling the theory to an irrelevant operator $\ocal$ at the lattice momentum $k_L$: $H = H_0 - \epsilon \, \ocal(k_L)$. It follows that $\dot {\vec P} = i [H,\vec P] = \epsilon \, \vec k_L \ocal(k_L)$.\footnote{In this case one finds that due to the projections ${\mathcal Q}$ in the definition of the memory matrix (\ref{eq:memory}), then $M_{PJ} \sim M_{PQ} \sim \epsilon^2$ are of the same order as $M_{PP}$, but our general statement above that $\left(M^{-1}\right)_{PP} \sim \epsilon^{-2} \gg 1$ dominates the inverse of the memory matrix remains true \cite{Hartnoll:2012rj}.}

With $\left(M^{-1}\right)_{PP}$ dominating the inverse memory matrix, the d.c. conductivities (\ref{eq:gam}) are
\be\label{eq:ppdominates}
\sigma = \rchi_{JP}^2 \left(M^{-1}\right)_{PP} \,, \qquad \overline \kappa =  \frac{1}{T} \rchi_{QP}^2 \left(M^{-1}\right)_{PP} \,.
\ee
Taking the ratio of the conductivities in (\ref{eq:ppdominates}), we obtain the advertised (\ref{eq:newWF})
\be\label{eq:newWF2}
\frac{\overline \k}{\s T} = \frac{1}{T^2} \frac{\rchi_{QP}^2}{\rchi_{JP}^2} \,.
\ee
In this formula for the ratio of conductivities the mechanism of momentum relaxation has cancelled out, leaving an expression in terms of purely thermodynamic quantities. In this sense (\ref{eq:newWF2}) captures a universality analogous to that of the
Wiedemann-Franz law. The value and indeed temperature dependence of this ratio will however depend on the low energy theory describing the system.

Fast relaxation of currents but slow relaxation of momentum is characteristic of hydrodynamic transport. For this reason we call these theories quasi-hydrodynamic. It has been suggested that a similar notion applies to highly correlated electron gases in semiconductor heterostructures \cite{ks1,ks2}. Indeed our formula (\ref{eq:newWF2}) can also be derived, in the relativistic case at least and now at sufficiently high temperatures, using standard hydrodynamics \cite{Hartnoll:2007ih}. We are, however, in a low temperature regime that is not that of conventional hydrodynamics. The memory matrix is the appropriate theoretical framework for us. See also \cite{varmao}.

An interesting result is obtained if we consider the thermal conductivity $\k = \overline \kappa - \alpha^2 T/\sigma$ in this case. Extending the dominance of $\left(M^{-1}\right)_{PP}$ in (\ref{eq:ppdominates}) to the thermoelectric conductivity $\a$, one finds that the two terms in the expression for $\k$ exactly cancel. We explain the simple physics of this cancellation at the end of the following subsection. The leading nonvanishing term in the thermal conductivity is a universal quantity computable in the low energy effective theory. This universal term does not benefit from the enhancement by $\left(M^{-1}\right)_{PP} \gg 1$. It follows that
\be\label{eq:extreme}
\frac{\kappa}{\sigma T} \ll 1 \,,
\ee
for these systems. A small Lorenz number therefore seems to be a characteristic, model-independent, feature of quasi-hydrodynamic non-Fermi liquid metals. This is perhaps the most dramatic of the kinematically-driven results we will find. It illustrates the limitations of a
weakly interacting intuition; the result (\ref{eq:extreme}) cannot be understood starting from a Fermi liquid and then perturbatively adding the effects of additional neutral degrees of freedom (more heat conduction) or additional inelastic scatterings (more heat relaxation).
In the Fermi liquid discussion of section \ref{sec:FL}, all relaxation was controlled by the matrix $\Gamma$ of density relaxation rates. In strongly interacting non-Fermi liquid metals, in contrast, momentum relaxes via a different mechanism to heat and electric current.

Many of the non-Fermi liquid theories that have been explicitly constructed have been based on a patch picture in momentum space, mimicking the low energy structure of a Fermi liquid, and will be considered in the following subsection. Typically these theories, however, are not under complete theoretical control. It seems to be an open question whether in the true IR of these theories there is one or many independently conserved momenta. On the other hand, controlled theories with only one conserved momentum and no patch structure at low energy abound in holographic approaches to metals. One upshot of our work is that the number of conserved momenta has direct experimental consequences. While some holographic phases of matter do exhibit intrinsic properties of Fermi surfaces such as logarithmic violation of the boundary scaling of the entanglement entropy \cite{Ogawa:2011bz, Huijse:2011ef} or low energy spectral weight at nonzero momentum \cite{Hartnoll:2012wm, Anantua:2012nj}, models considered thus far do not show the large number of conserved densities and momenta associated to a patchlike low energy theory. It is interesting that essential properties of a Fermi surface that occur together in a weakly coupled language -- e.g. logarithmic entanglement, finite momentum spectral weight and an infinite number of conserved momenta -- can be dissociated in non-quasiparticle based theories.

\subsection{Non-Fermi liquids with patchwise conserved momenta}
\label{sec:nflmany}

If the low energy degrees of freedom of the non-Fermi liquid decouple into patches in momentum space, then there will be many conserved momenta. The difference with the Fermi liquid case will be that the individual patch theories will not be free. Thus the only almost conserved quantity in each patch will be the momentum $\vec P_\theta$. The ratio of conductivities (\ref{eq:newWF2}) becomes
\be\label{eq:complicatedWF2}
\frac{\overline \kappa}{\sigma T} = \frac{1}{T^2} \frac{\int d\theta d\theta' \rchi_{Q_\theta P_\theta} \left(M^{-1}\right)_{P_\theta P_{\theta'}} \rchi_{P_{\theta'} Q_{\theta'}}}{\int d\theta d\theta' \rchi_{J_\theta P_\theta} \left(M^{-1}\right)_{P_\theta P_{\theta'}} \rchi_{P_{\theta'} J_{\theta'}}} \,.
\ee
Part of the assumption of decoupled patches here is that the interpach susceptibilities vanish to leading order at low energies. This requires that the Landau interpatch interactions are irrelevant in these non-Fermi liquids or that they can be `diagonalized' to give decoupled patchwise theories. We see immediately that the patchwise susceptibilities entering the above formula are distinct from those appearing in the Fermi liquid case (\ref{eq:derivewf}). Transport in non-Fermi liquids is controlled by a different underlying kinematical structure.

In appendix \ref{sec:example} we describe the Ising-nematic quantum phase transition in two dimensional metals as an example of theory \cite{Metlitski:2010pd} with a strongly interacting patchwise description at low energies. The almost conserved quantities are the momenta in each patch.

The expression (\ref{eq:complicatedWF2}) for the ratio of conductivities is less universal than our result (\ref{eq:newWF2}) for quasi-hydrodynamic metals. Assuming that the irrelevant interpatch scattering is controlled by one scale to leading order at low energies, 
we can write $\left(M^{-1}\right)_{P_\theta P_{\theta'}} = \lambda \, F(\theta,\theta')$. Here $\lambda$ is a rate of momentum relaxation whereas $F(\theta,\theta')$ is a dimensionless `kinematic' function of pairs of points on the Fermi surface that does not contain another scale. Then we can write, at least in terms of extracting the temperature dependence of the ratio,
\be\label{eq:schematicWF2}
\frac{{\overline \kappa}}{\sigma T} \sim \frac{1}{T^2} \left\langle \frac{\rchi_{PQ}^2}{\rchi_{PJ}^2} \right\rangle
\sim \frac{\kappa}{\sigma T} \,.
\ee
The angled brackets here denote a schematic average over the Fermi surface in which the relaxation rate $\lambda$ has cancelled out.

To obtain the second relation in (\ref{eq:schematicWF2}), we can first explain why the cancellation we found above for quasi-hydrodynamic metals in the computation of $\kappa = {\overline \kappa} - \alpha^2 T/\sigma$ does not occur here. Recall that $\kappa$ is defined as the heat conductivity at vanishing electric current. Because $\rchi_{JP} \neq 0$ for the metallic states we are considering, the no-current boundary condition requires that the total momentum also vanish. For the quasi-hydrodynamic metals, the total momentum was the only conserved quantity. Therefore, in states with vanishing total momentum, the heat current can relax and heat conduction is universal. In the metals with patchwise conserved momenta, however, the vanishing of the total current does not imply that all of the patch momenta must independently vanish. The nonvanishing patch momenta will then not allow the total heat current to relax within the momentum-conserving low energy effective theory. This can be seen explicitly by verifying that because of the integrals over patches on the Fermi surface in expressions like (\ref{eq:complicatedWF2}), the $\left(M^{-1}\right)_{P_\theta P_{\theta'}}$ dependence no longer cancels out in $\kappa = {\overline \kappa} - \alpha^2 T/\sigma$. In addition to the absence of a cancellation, we can argue that $\kappa \sim {\overline \kappa}$ have the same temperature scaling: In these non-Fermi liquids where only the patchwise momenta are conserved, only $\rchi_{PJ}$ and $\rchi_{PQ}$ appear in the conductivities. It is then easily seen that, unlike in the Fermi liquid case,
$\alpha^2 T/\sigma$ and ${\overline \kappa}$ contain exactly the same susceptibilities and therefore have the same temperature dependence. Thus we finally obtain the second relation in (\ref{eq:schematicWF2}).

Universal heat conduction has also been obtained theoretically in the past by dividing the system into two sets of modes. Among the first set of modes, interactions degrade all currents while the momentum can be transferred to the second set of modes. The second set of modes is then assumed to dissipate momentum very quickly. In these circumstances a universal heat conduction can be associated to the first set of modes. See e.g. \cite{fritz, smv}.

\section{Non-Fermi liquids with (some) long-lived quasiparticles}
\label{sec:long}

In many experimental systems believed to be close to quantum critical points, the order parameter carries a finite wavevector. For instance in metallic spin and charge density wave transitions. In such cases, fluctuations of the order parameter are most efficient at scattering low energy fermions in the vicinity of hot spots or hot lines on the Fermi surface. The hot loci are connected in momentum space by an ordering wavevector while the remaining patches on the Fermi surface are referred to as `cold'. The cold fermions can in turn scatter off excitations at the hot loci \cite{Hartnoll:2011ic}. In this section we discuss the case in which at least some long-lived `cold' quasiparticles survive in the regime showing non-Fermi liquid transport.

The possibility of long-lived quasiparticles coupled to strongly interacting critical excitations also arises in systems with multiple bands. One band can become `hot' while the other `cold' bands retain a quasiparticle character. Evidence for this phenomenon can be seen in quantum oscillation experiments in e.g. Sr$_3$Ru$_2$O$_7$ \cite{dvaSr}. Interband scattering has the potential to lead to unconventional lifetimes for the stable quasiparticles.

In terms of our effective field theory approach to quantum transport, in the cases where quasiparticle and non-quasiparticle excitations coexist,
one must consider separately the strongly interacting patches at or close to the hot loci and the quasiparticle patches away from the hot loci.
For the long-lived cold fermions, we can apply the memory matrix method much as for Fermi liquids.
As we noted in appendix \ref{sec:strong}, in spin and charge density wave transitions in two space dimensions, the theory of the hot loci involves strong scattering with a nonzero momentum transfer \cite{Hartnoll:2011ic}. Momentum is therefore not approximately conserved and the memory matrix method is not applicable to the hot excitations. On the other hand, this fact also suggests that once momentum is transferred from the cold patches to the hot loci via weak interpatch scattering, then it is relaxed very quickly.

The interplay of hot and cold patches has immediate consequences for transport. An important likelihood is that any anomalously large resistivity from the hot patches will be short circuited by the cold fermions \cite{hr}. This is not necessarily the case in all temperature regimes \cite{ros}. We proceed to consider situations in which the cold fermions dominate the transport, but acquire non-Fermi liquid lifetimes through scattering off hot excitations, while remaining well-defined quasiparticles.
We will see that such scenarios allow the Wiedemann-Franz law to coexist at low temperatures with non-Fermi liquid transport.

\subsection{Linear in temperature resistivity and Wiedemann-Franz}
\label{sec:linear}

In section \ref{sec:data} below we will see that several materials exhibit non-Fermi liquid transport while simultaneously satisfying the
Wiedemann-Franz law. In this section we show that the coexistence of the Wiedemann-Franz law with unconventional transport places strong constraints on the dissipative process controlling the transport.

It is well known that for temperatures above the Debye temperature but below the Fermi energy, scattering of electrons by phonons leads to a linear in temperature resistivity and also to the Wiedemann-Franz law \cite{ziman}. There are four important facts in this temperature range that allow these two phenomena to coexist: (i) the rate at which the electronic quasiparticles lose momentum to the phonons is given by $\Gamma \sim T$, (ii) the phonons lose momentum via Umklapp scattering at a faster rate $\Gamma_U \gg \Gamma$, (iii) the electronic quasiparticles remain long lived and (iv) because the average energy of the fermions involved in the scattering is much larger than that of the phonons (whose energy is bounded above by the Debye energy), the electrons effectively experience elastic scattering. It is clear that points (iii) and (iv) allow the derivation of Wiedemann-Franz in section \ref{sec:FL} to go through. Points (i) and (ii) are additionally required because they ensure that electron scattering immediately leads to momentum dissipation, allowing currents to relax. Point (ii) is furthermore related to the fact that the electronic contribution to thermal conduction will be much bigger than the phonon contribution.

We wish to know what general types of emergent collective low energy excitations could satisfy conditions (i) to (iv). The experimental case has been made that some linear in temperature resistivity is due to scattering of well-defined quasiparticles by critical modes that are classicalized by being above their effective Debye temperature \cite{universal}. To substantiate this picture a framework for very low or even vanishing effective Debye scales is necessary.

We can take a general approach. Suppose we are given a critical bosonic mode $\ocal$ with a retarded Green's function $D^R(\omega,k)$. Let the scattering of the cold fermions $\psi$ with the bosonic mode be described by the coupling $S_\text{int} = \lambda \int dt d^dx \psi^\dagger \psi \ocal$. To describe scattering by hot fermions, we can imagine that the operator $\ocal$ is a fermion bilinear.
We assume that this coupling can be treated perturbatively at low energies, consistent with the survival of the cold fermions as quasiparticles. This kind of coupling between cold fermions and hot operators was considered in \cite{Hartnoll:2011ic} for the metallic spin density wave transition. If we assume that the hot excitations are able to dissipate momentum very efficiently, as seems to be the case for the spin density wave transition in two dimensions \cite{Hartnoll:2011ic}, then the transport will be dominated by cold quasiparticles. This transport will be controlled by the decay rate
\be\label{eq:Sigma}
\text{Im} \Sigma(\w,p) = \lambda^2 \int \frac{d^dk}{(2\pi)^d} \frac{d\Omega}{\pi}
\text{Im} G^R(\w-\Omega,p-k) \, \text{Im} D^R(\Omega,k) \frac{f_\text{FD}(\w-\Omega) f_\text{BE}(\Omega)}{f_\text{FD}(\w)} \,.
\ee
This formula is essentially Fermi's golden rule allowing for decay into non-quasiparticle modes and is easily derived using standard thermal field theory methods. Here $f_\text{FD}$ and $f_\text{BE}$ are the Fermi-Dirac and Bose-Einstein distributions, respectively. In general we should allow for the coupling $\lambda$ to be $k$ dependent. For the free fermion
\be\label{eq:delta}
\text{Im} G^R(\omega,k) = \pi \, \delta(\omega - \epsilon(k)) \,,
\ee
where $\epsilon(k)$ is the free fermion dispersion and vanishes on the Fermi surface. We can now ask: for what Green's functions $D^R(\omega,k)$ is a linear in temperature resistivity obtained? When is this scattering sufficiently elastic that the Wiedemann-Franz law is true?

Elasticity will hold when the energy transfer is much less than the typical energy of the cold electrons, namely when $\Omega \ll \w \sim T$ throughout the integrand in (\ref{eq:Sigma}). The fermion energy $\w$ is measured from the chemical potential. From (\ref{eq:Sigma}), expanding the factor of $f_\text{BE} \approx T/\Omega$ and noting the cancellation of the two factors of $f_\text{FD}$ in this regime of energies, we obtain
\be\label{eq:linT}
\text{Im} \Sigma(\w,p) = \lambda^2 T \int \frac{d^dk \, d\Omega}{(2\pi)^d} \, \frac{\text{Im} D^R(\Omega,k)}{\Omega} \delta(\Omega - \omega + \epsilon(p-k)) \,.
\ee
If the bosonic spectral weight $\text{Im} D^R$ does not have a strong temperature dependence, and dies off sufficiently quickly at large energies so that there is no contribution to the integral from $\Omega \sim T$, then we immediately obtain a linear in temperature scattering rate. For instance, for Debye phonons,
\be\label{eq:phonon}
\text{Im} D_\text{phon.}^R(\Omega,k) = \delta(\Omega^2 - c_s^2 k^2)\,.
\ee
The lattice cutoff on the momentum $k < k_L$ therefore implies that $|\Omega| < c_s k_L$. It follows that if the temperature is above the Debye temperature, $c_s k_L \ll T$, we consistently find elastic scattering with a linear in temperature decay rate. This is of course a textbook result for high temperature scattering of electrons by phonons.

Given that the boson in the cases of interest is emerging from a strongly correlated sector, we do not expect its spectral weight to take the free form (\ref{eq:phonon}). We will call a bosonic mode with a spectral weight $\text{Im} D^R(\Omega,k)$ such that there exists an effective Debye scale $\Omega_D$, so that for $T \gg \Omega_D$ equation (\ref{eq:linT}) holds, a `generalized phonon'.

If classicalized phonons provide one canonical way to obtain a linear in temperature scattering rate via a bosonic mode, the other prototypical framework is the marginal Fermi liquid \cite{varma}. There the bosonic spectral weight is postulated to have the form
\be
\text{Im} D_\text{MFL}^R(\Omega,k) \sim
\begin{cases}
\Omega/T & \Omega \lesssim T \\
\text{sgn}(\Omega) & \Omega \gtrsim T
\end{cases} \,,
\ee
over some large range of momentum. Setting $\Omega = T \hat \Omega$ and $\omega = T \hat \omega$ in the formula for the lifetime (\ref{eq:Sigma}) and using (\ref{eq:delta}) we again obtain a linear in temperature relaxation rate. This is a completely different regime from the phonon scattering. The energy transferred to the bosonic mode saturates the temperature scale and hence the scattering is manifestly not elastic. Independently of how efficiently the boson can lose its momentum and the extent to which the boson contributes to thermal transport, the mechanism is not compatible with the Wiedemann-Franz law. This is because inelasticity means that we have lost the simple relation between current and heat relaxation that played a key role the derivation of the law in section \ref{sec:FL}.

The Wiedemann-Franz law therefore is a diagnostic that can differentiate marginal Fermi liquid-like from phonon-like linear in temperature relaxation rates.

As is well known, the fermion relaxation rate and the transport relaxation rate can differ if small-angle scattering dominates transport. At  temperatures above the effective Debye temperature of the `generalized phonon' mode, typical scatterings involve a momentum transfer of order the lattice momentum. These large momentum scatterings relax currents and momenta at the same rate as the fermions relax. It is therefore sufficient to consider the quasiparticle decay rate (\ref{eq:Sigma}) to obtain the d.c. resistivity.

Several materials showing a linear in temperature resistivity at low temperatures also show a specific heat with a temperature dependence of $c \sim - T \log T$, e.g. \cite{stewart, rost}. A free phonon, with spectral weight (\ref{eq:phonon}), is well known to contribute a constant specific heat, contradicting the observations.\footnote{This point was emphasized to us by Andy Mackenzie.} However, we do not expect the critical mode to be a free boson. The contribution of a `generalized phonon' with retarded Green's function $D^R(\omega,k)$ to the specific heat is computed in appendix \ref{sec:c}. It is found that the simplest generalized phonons of (\ref{eq:genp}) again require an (unobserved) constant contribution to the specific heat at low temperatures, as we might have anticipated from energy equipartition. This fact motivates experimental probes that are able to resolve a small constant contribution to the specific heat at low temperatures in materials showing a linear in $T$ resistivity with Wiedemann-Franz simultaneously satisfied.

\section{Revisiting the experiments}
\label{sec:data}

Following the discussions above of various flavors of non-Fermi liquids with and without quasiparticles, it is instructive to revisit the experimental results.

We primarily discuss the heavy fermions YbRh$_2$Si$_2$, CeCoIn$_5$, CeRhIn$_5$ and the ruthenate Sr$_3$Ru$_2$O$_7$. 
These materials exhibit non-Fermi liquid behavior at, or in the vicinity of, a metallic critical point. In applying our considerations to these materials we are implicitly assuming that they show well-defined Drude peaks. In the absence of magnetic fields or pressure, the heavy fermion optical conductivities have been studied in e.g. \cite{optical1, optical2}. A sharp Drude peak was observed over temperature ranges showing non-Fermi liquid transport. Our discussion ignores the direct effects of magnetic fields on transport. Magnetic fields can be incorporated into the memory matrix formalism and have a strong effect on the d.c. conductivities if they dominate over other sources of momentum relaxation \cite{Hartnoll:2007ih}. The ratio of electric and thermal Hall conductivities in CeCoIn$_5$ was studied in \cite{ong}.

\begin{figure}[h]
\begin{center}
\includegraphics[height=180pt]{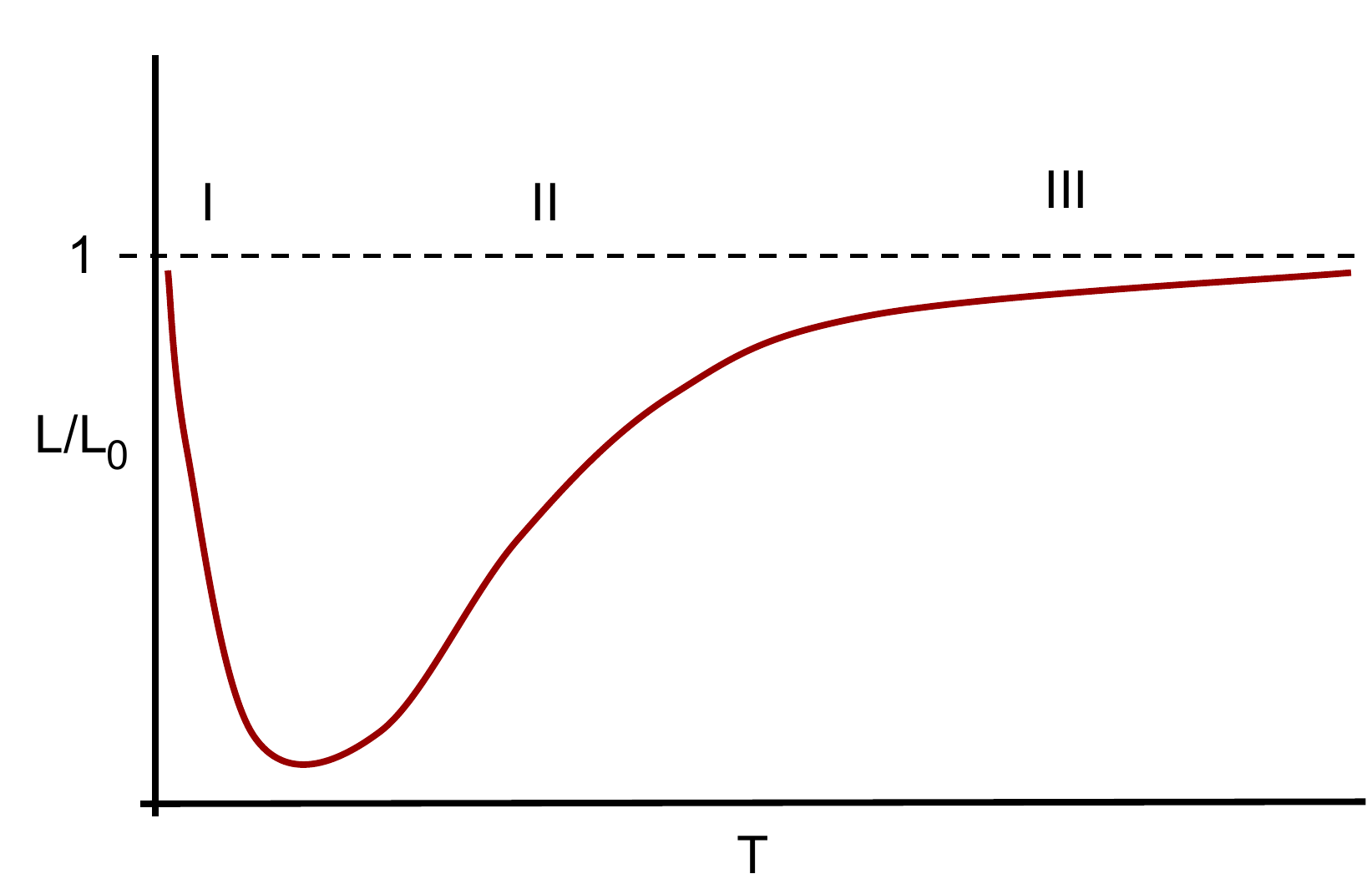}
\caption{Ratio of conductivities as a function of temperature in conventional metals and in certain non-Fermi liquids. Wiedemann-Franz is satisfied in regions I and III, but not in the intermediate region II.
In conventional metals the crossover from II to III occurs at the Debye temperature. The non-Fermi liquids under discussion exhibit the same structure at very low temperatures.
\label{fig:LL0}}
\end{center}
\end{figure}

A plot of the ratio of the electronic contribution to the conductivities $L/L_0 \equiv \kappa/\sigma T \, \times \, 3/\pi^2$ as a function of low temperature in CeCoIn$_5$, CeRhIn$_5$ and Sr$_3$Ru$_2$O$_7$ \cite{wf3,caxis,wf5,wf22} shows the same shape as is observed in conventional metals due to scattering off phonons and impurities \cite{ziman}. That is, there are three regimes: At the highest temperatures, Wiedemann-Franz is obeyed due to scattering by phonons above their Debye temperature, as we recalled in the previous section. Below the Debye temperature, Wiedemann-Franz is violated due to the onset of inelastic scattering by phonons. At the lowest temperatures, Wiedemann-Franz is again recovered due to elastic impurity scattering dominating. We sketch this situation in figure \ref{fig:LL0}. For the non-Fermi liquids, however, the temperature scales at which these transitions occur are well below the Debye scale of the metals and therefore are not related to actual phonon scattering.

The relevant experimental information is summarized in table \ref{tab:mat} below. In this table we see that the materials fall into two distinct classes.
\vspace{0.3cm}

\begin{table}[h]
\begin{center}
\centerline{\begin{tabular}{|l|c|c|c|}
\hline
Material & \specialcell{region I: WF \\ very low $T$} & \specialcell{region II: {\xcancel{WF}} \\ intermediate low $T$} & \specialcell{region III: WF \\ higher low $T$} \\
\hline\hline
CeCoIn$_5$ (a-axis, critical field) \cite{wf3,caxis} & yes & crossover to $\rho \sim T^{3/2}$  & $\rho \sim T$ \\
\hline
CeRhIn$_5$ (non-critical field, AFM) \cite{wf22} & yes & $\rho \sim T^2$ & $\rho \sim T$ \\
\hline
Sr$_3$Ru$_2$O$_7$ (non-critical field) \cite{wf5,universal,rost} & yes & $\rho \sim T^2$ & $\rho \sim T$ \\
\hline
Sr$_3$Ru$_2$O$_7$ (critical field) \cite{wf5,universal,rost} & yes & $\rho \sim T$ & $\rho \sim T$ \\
\hline
YbRh$_2$Si$_2$ (critical field) \cite{gegen, con1,con1a,con2} & disputed & $\rho \sim T$ & no measurements \\
\hline
CeCoIn$_5$ (c-axis, critical field) \cite{caxis} & no & $\rho \sim T$ & no measurements \\
\hline
\end{tabular}}
\vspace{0.3cm}
\caption{Schematic electrical resistivity in the three low temperature regimes of figure \ref{fig:LL0}, characterized by whether or not Wiedemann-Franz holds, for several non-Fermi liquid materials. In the description of the materials, `critical field' means that the material has been tuned to a quantum critical point by an external magnetic field. `No measurements' means that a higher temperature region III with Wiedemann-Franz satisfied is not found at the temperatures currently reported. The materials can be divided into those in which the linear resistivity is observed simultaneously with Wiedemann-Franz holding and those in which it is not. This suggests a division into non-Fermi liquids that do not and do, respectively, admit a quasiparticle description of transport, as we discuss in the text. Critical Sr$_3$Ru$_2$O$_7$ is a confusing exception here, as we also discuss in the text. \label{tab:mat}}
\end{center}
\vspace{-1cm}
\end{table}

\subsection*{Non-critical Sr$_3$Ru$_2$O$_7$, CeRhIn$_5$ and critical a-axis CeCoIn$_5$}

In table \ref{tab:mat}, three cases show a crossover from Wiedemann-Franz together with linear in temperature resistivity above some temperature to violation of Wiedemann-Franz together with a resistivity scaling like $\rho \sim T^{x}$, with $x>1$, below that temperature (Sr$_3$Ru$_2$O$_7$ and CeRhIn$_5$ away from criticality and a-axis transport in CeCoIn$_5$ at criticality). This is exactly what one expects for scattering of well-defined quasiparticles by a `generalized phonon' mode going through its effective `Debye temperature'. Restoration of Wiedemann-Franz at the lowest temperatures in these materials is then also to be expected because the electronic quasiparticle has not been destroyed.

The behavior of these three materials is consistent with the kinematical framework of quantum criticality combined with long-lived particles described in section \ref{sec:long} above.
 
 \subsection*{Critical YbRh$_2$Si$_2$, c-axis CeCoIn$_5$ and Sr$_3$Ru$_2$O$_7$}

In critical YbRh$_2$Si$_2$ and Sr$_3$Ru$_2$O$_7$ and c-axis transport in critical CeCoIn$_5$, in contrast, a linear in temperature resistivity is found in temperature regimes in which Wiedemann-Franz is not satisfied. This transport cannot be explained by a Debye scale collapsing to zero at the critical point. The existing data for YbRh$_2$Si$_2$ and c-axis CeCoIn$_5$ in this regime is compatible with strongly interacting transport without quasiparticles of the sort described in section \ref{sec:nfl} above. If this is the case we can expect formulae like (\ref{eq:schematicWF2}) to hold and describe the violation of the Wiedemann-Franz law. In both these materials, over the low temperature range showing a linear resistivity, the Lorenz ratio has a simple temperature dependence. It appears either almost constant or weakly linear in temperature \cite{caxis, con1, con1a}. This is suggestive of the notion that these dependences could be obtained from a ratio of susceptibilities such as (\ref{eq:schematicWF2}). If this is true, we can predict that $L/L_0$ will continue to be constant or linear at larger temperatures, while the linear in temperature resistivity holds, and Wiedemann-Franz will not be recovered.\footnote{As this paper was completed, measurements of the Lorenz ratio in YbAgGe were announced \cite{newWF}. At the critical magnetic field, these showed $L/L_0$ depending linearly on temperature over a range of low temperatures, violating Wiedemann-Franz and furthermore crossing straight through the Wiedemann-Franz value. If the thermal transport is not dominated by phonons -- see the caveats described in \cite{newWF} -- then these results are potentially a striking realization of strongly correlated transport as we have characterized it in section \ref{sec:nfl}: in the critical regime, showing a linear in temperature electrical resistivity, the Wiedemann-Franz value for the ratio of conductivities does not appear as a useful reference point for understanding the transport.} It would be interesting to measure $\overline \kappa$ in these materials, where any universality may be expected to be more pronounced. In fact, the temperature scaling $\k \sim {\overline \k}$ without strict equality (recall that $\k = {\overline \k}$ for quasiparticles) is a diagnostic for strongly correlated transport with patchwise conserved momenta. Confirmation of $\k \sim {\overline \k}$ would rule out a quasiparticle based marginal Fermi liquid-type explanation for the linear in temperature resistivity in these materials.

The possible restoration of Wiedemann-Franz as $T \to 0$ in YbRh$_2$Si$_2$ at the critical field \cite{con1a,con2} would indicate an energy scale that is not collapsing at the critical point.

It is hard to understand, however, how in Sr$_3$Ru$_2$O$_7$ tuned to criticality Wiedemann-Franz can be restored at higher temperatures without even the slope in the linear resistivity changing. This is because the linear resistivity would seem to be necessarily due to different scattering mechanisms in the two regimes (i.e. with and without Wiedemann-Franz holding). The observations of thermal conductivity in Sr$_3$Ru$_2$O$_7$ in \cite{wf5} predated the more precise measurements, on purer samples, of electrical conductivity in \cite{rost, universal}. Perhaps the thermal conductivity in the critical regime should be revisited. Our discussion suggests that either the Wiedemann-Franz law holds down to low temperatures at the critical magnetic field, or there is a change in the temperature dependence of the resistivity.

\subsection*{Cuprates}

The existing experimental discussion of Wiedemann-Franz in the cuprates is mostly concerned with very low temperatures in which the electrical resistivity is either constant or increasing as the temperature is lowered, the latter case due to the proximity of localized phases. These results do not probe the linear in temperature resistivity.

The observation of the Wiedemann-Franz law in the overdoped cuprates \cite{wf2} is consistent with well-defined quasiparticles scattering elastically off impurities at zero temperature. Several explanations have been given for the reported violation of Wiedemann-Franz in underdoped and optimally doped cuprates \cite{nwf1,nwf2,nwf3} at very low temperatures (suppressing superconductivity as necessary). We limit ourselves to the comment that, if violation of Wiedemann-Franz is due to strong interactions and no quasiparticles, then there is no reason to prefer $L/L_0$ larger \cite{nwf1, nwf2} or smaller \cite{nwf3} than unity. As we have repeatedly stressed throughout, $L_0$ is not a legitimate reference point in strongly interacting circumstances that are governed by different kinematics than those underlying the Wiedemann-Franz law.

The Hall Lorenz ratio has been measured in optimally doped YBCO at temperatures with a linear resistivity \cite{zhang}. The use of Hall conductivities removes the phonon contribution to heat transport. Wiedemann-Franz does not hold, and the Lorenz ratio shows a clean linear dependence on temperature. This behavior is consistent with the Lorenz ratio being given by a ratio of thermodynamic susceptibilities like (\ref{eq:schematicWF2}) in the absence of quasiparticles.

\section{Discussion: The two faces of quantum criticality}

In this work we have focussed on the kinematics of non-Fermi liquid transport. The starting point for any discussion of transport in a metal, in particular if strong interactions are involved, needs to be: What are the almost-conserved quantities governing the low energy dynamics?

The focus on almost-conserved quantities split our discussion of non-Fermi liquid transport into two cases. In the first case of `total quantum criticality', all quasiparticles are destroyed by strong interactions at low energies. Therefore all transport processes occur via strongly interacting modes and the only almost conserved quantities are momenta. In the second case of `backseat quantum criticality', only a subset of the degrees of freedom are strongly interacting at low energies. The long lived degrees of freedom can acquire unconventional lifetimes due to scattering off the critical modes, but the critical modes themselves do not participate directly in transport as they e.g. relax their momentum too quickly.

In a `total quantum criticality' scenario the Wiedemann-Franz law is completely off the map. Because the kinematics of almost-conserved quantities is not related to that of a Fermi liquid, one cannot understand these cases by starting with a Fermi liquid and then imagining adding perturbatively additional neutral heat carriers (to increase the thermal conductivity) or additional inelastic scattering (to increase thermal resistivity).

In our discussion of totally quantum critical transport it was interesting to distinguish the theoretically natural ratio
$\overline \kappa/\sigma T$, involving the thermal conductivity at vanishing electric field $\overline \kappa$, from the usual $\kappa/\sigma T$. For long-lived fermionic quasiparticles $\kappa = {\overline \kappa} \gg \a^2 T/\sigma$, for hydrodynamic non-Fermi liquids $\kappa \ll {\overline \kappa}$, while for non-Fermi liquids with patchwise conserved momenta $\kappa \sim {\overline \kappa} \sim \a^2 T/\sigma$. The ratio $\overline \kappa/\sigma T$ in totally critical non-Fermi liquids
was universally related to a ratio of thermodynamic susceptibilities in (\ref{eq:newWF2}) or, slightly less powerfully, (\ref{eq:schematicWF2}). Perhaps these susceptibilities can be independently extracted experimentally. We noted in section \ref{sec:data} that the ratio of conductivities does show a fairly regular temperature dependence in the relevant temperature regimes of the candidate totally quantum critical materials YbRh$_2$Si$_2$, c-axis CeCoIn$_5$ and cuprates.

`Backseat quantum criticality' is compatible with the Wiedemann-Franz law if certain circumstances hold. In particular, a linear in temperature resistivity can coexist with the Wiedemann-Franz law if it is caused by scattering off a `generalized phonon' mode above its effective `Debye temperature'. Crucially, in these cases, Wiedemann-Franz is expected to hold {\it above} a certain low temperature. This behavior has been observed in CeRhIn$_5$ and Sr$_3$Ru$_2$O$_7$ away from the critical magnetic fields and in a-axis transport in CeCoIn$_5$ at the critical field. The linear in temperature resistivity observed in these materials would seem therefore to be of a fundamentally different nature to that in the materials mentioned in the previous paragraph, where the linear resistivity did not coexist with the Wiedemann-Franz law. The Wiedemann-Franz law is therefore an interesting diagnostic of quantum critical physics not just at the lowest possible temperature scales, but also higher temperatures.

\section*{Acknowledgements}

In writing this paper we have benefitted greatly from discussions and correspondence with  Diego Hofman, Nigel Hussey, Andy Mackenzie, John McGreevy, Max Metlitski, Joe Polchinski, Jean-Phillipe Reid, Subir Sachdev, Todadri Senthil, Brian Swingle, Louis Taillefer, William Witczak-Krempa, Jan Zaanen and especially Steve Kivelson. SAH is partially funded by a DOE Early Career Award and by a Sloan fellowship. RM is supported by a Gerhard Casper Stanford Graduate Fellowship. MB is supported by the Simons Foundation.

\appendix

\section{Strong momentum relaxation}
\label{sec:strong}

This appendix describes circumstances in which our discussion of transport in strongly interacting systems -- organized around the existence of almost conversed momentum or momenta in the effective low energy theory -- does not apply.

Strong momentum-violating interactions can manifest themselves in two ways. The first occurs when an emergent particle-hole symmetry in the low energy theory results in susceptibilities such as $\rchi_{JP}$ going to zero. Then the current can relax quickly without being `dragged along' by the emergent almost-conserved momentum. The second possibility is that the effects of scattering become so strong that the momentum relaxation rate $\Gamma \to \infty$. These two scenarios are distinguished by the fact that in the first case the spectral weight of the Drude peak vanishes while in the second the weight is conserved but the peak becomes infinitely broad. We briefly discuss these two cases in turn.

Emergent particle-hole symmetry arises in the metallic spin density wave quantum phase transition in two dimensions \cite{Metlitski:2010vm} and in certain theories of continuous Mott transitions \cite{senthil1, senthil2}. In the spin density wave case, umklapp scattering by finite momentum critical modes connecting the hot spots contributes a universal critical conductivity \cite{Hartnoll:2011ic}. We shall return to spin density wave transistions in section \ref{sec:long}, as transport is dominated by excitations away from the hot spots. For the Mott transitions, universal charge transport is obtained from the particle-hole symmetric dynamics of fluctuations about the half-filled state \cite{senthil1, senthil2}. Particle-hole symmetry is insufficient to obtain universal heat transport, as $\rchi_{QP}$ will typically not vanish. Therefore heat transport will be tied to the non-universal fate of momentum conservation. We will qualify this statement in section \ref{sec:nflmany}, as there are circumstances in which $\kappa$ is universal even while $\overline \kappa$ is not.

A divergent momentum relaxation rate in a theory without disorder requires that an umklapp-like operator become relevant in the low energy theory. A localization transition driven via this mechanism has recently been realized in a holographic context \cite{Donos:2012js}.

\section{Patchwise conserved momenta: Ising-nematic example}
\label{sec:example}

In this appendix we consider the theory of the Ising-nematic quantum phase transition in two dimensional metals, developed recently in \cite{Metlitski:2010pd}, as an example of a strongly interacting theory with a patchwise description. Earlier work on this model includes \cite{Polchinski:1993ii, Altshuler:1994zz, sslee}. While the loop expansion is ultimately not controlled in this model, it will serve to illustrate the different ways in which various quantities relax. The theory is obtained by zooming in on a pair of antipodal patches of a Fermi surface and maintaining a scaling regime where interactions with a collective boson are strong. Because the fermions only interact efficiently with the boson when the boson momentum is parallel to the Fermi surface, due to the Fermi surface curvature, to capture this process it is necessary to `thicken' the patches. Thus, unlike in the Fermi liquid case, the patchwise theory describes propagation in two space dimensions \cite{Metlitski:2010pd}. The effective patchwise low energy theory of \cite{Metlitski:2010pd} describes two antipodal fermions $\psi_s$, with $s=\pm$, interacting with a gapless boson $\phi$
\be\label{eq:lag}
{\mathcal{L}} = \sum_{s=\pm} \psi^\dagger_s \left(i\eta \pa_t + i s \pa_x + \pa_y^2 \right) \psi_s
- \sum_{s = \pm}
\lambda_s \phi \psi^\dagger_s \psi_s + \frac{N}{2 e^2} \left[\eta' \left(\pa_t \phi \right)^2 - \left(\pa_y \phi \right)^2 \right] \,.
\ee
The spin flavor index running from $1$ to $N$ has been suppressed. In the Ising-nematic model $\lambda_+ = \lambda_-$. This theory can also describe a spin liquid when $\lambda_+ = - \lambda_-$. The renormalization group (RG) scaling that one considers for this model is: $\pa_x \to b^2 \pa_x, \pa_y \to b \pa_y, \pa_t \to b^3 \pa_t$ together with $\psi \to b^2 \psi, \phi \to b^2 \phi$. The couplings $\lambda_\pm$ and $e$ are dimensionless under this scaling. We see that both of the time derivative terms we have included in the above Lagrangian are in fact irrelevant under this scaling. Thus we should consider $\eta, \eta' \to 0$. Nontrivial frequency dependence compatible with the RG scaling will be generated radiatively \cite{Metlitski:2010pd}. We have included the time derivative terms above in order to be able to easily define tree level operators for e.g. the momentum. The fact that the classical frequency dependence will be swamped at low energies by terms generated in the RG flow tells us that the momentum operators we will shortly define in fact undergo very significant vertex corrections, which we will ignore. The comments that follow should be taken at the level of a qualitative discussion of momentum versus current relaxation in these theories.

We proceed to compute the time derivatives of currents and momenta in the patch theory (\ref{eq:lag}). Our modest goal here is to use these expressions as an explicit example of how patchwise momenta but not currents are conserved in a strongly interacting theory. Given the Lagrangian (\ref{eq:lag}) we can write down the patch momenta and current densities using the Noether procedure and the equations of motion
\begin{align}
\vec p &  =  \sum_s \frac{i \eta}{2} \, \left(\vec \nabla \psi^\dagger_s \, \psi_s - \psi^\dagger_s \vec \nabla \psi_s  \right) + \frac{N}{e^2} \eta' \, \pa_\tau \phi \vec \nabla \phi \,, \\
j_x &  =  \sum_s s \psi^\dagger_s \psi_s \,, \\
j_y & =  i \sum_s \left(\pa_y \psi^\dagger_s \, \psi_s - \psi^\dagger_s \pa_y \psi_s \right) \,, 
\end{align}
\begin{align}
q_x & =  \frac{-1}{2\eta} \sum_s s
\left( \psi^\dagger_s (i s\partial_x + \partial_y^2) \psi_s + \textrm{c.c}  \right)
+ \frac{1}{\eta}\sum_s \lambda_s\, \phi\, \psi_s^\dagger \psi_s \,, \\
q_y & =  - \frac{1}{\eta}\sum_s \left(  \pa_y \psi_s^\dagger (-\pa_x + i \pa_y^2)\psi +  \textrm{c.c} \right)
- \frac{N}{e^2} \pa_y \phi\, \dot{\phi} \nonumber \\
&+ \frac{1}{\eta} \sum_s \lambda_s \phi \left( i \pa_y \psi_s^\dagger\psi_s + \textrm{c.c}\right)\,,
\end{align}
and the Hamiltonian density is
\be
h = \frac{N}{2 e^2} \left[\eta' \left(\pa_\t \phi \right)^2 + \left(\pa_y \phi \right)^2 \right] + \sum_s \left( \frac{is}{2} \left[\pa_x \psi_s^\dagger \psi_s - \psi_s^\dagger \pa_x\psi_s \right] +
\pa_y \psi_s^\dagger \pa_y \psi_s +
\lambda_s \phi \psi_s^\dagger \psi_s \right)\,.
\ee
The total patch quantities $\{\vec P, \vec J, \vec Q,H \}$ are obtained by integrating the densities over space. This allows us to compute the time derivatives
\begin{align}
i\dot {\vec P} &=  [P,\vec H] = 0 \,, \\
i\dot J_x &=  [J_x,H] = 0  \,, \\
i\dot J_y &=  [J_y,H] = \frac{-2i}{\eta}\sum_s \lambda_s \int d^2x\, \partial_y \phi\, \psi_s^\dagger \psi_s\,, \\
i\dot Q_x &= [Q_x,H] = \frac{i}{\eta} \sum_s \lambda_s \int d^2x \, \dot{\phi}\psi_s^\dagger \psi_s \,,
\end{align}
and
\begin{align}
i\dot Q_y &=  \sum_s \int d^2x \, \left( \left[-\frac{i\lambda_s}{\eta'}\phi + \frac{i\lambda_s^2}{\eta^2}\phi^2 \right]\partial_y (\psi^\dagger \psi)
+ \frac{i\lambda_s \phi}{\eta^2} \left[ \left(\partial_y^3 \psi^\dagger \psi - \partial_y \psi^\dagger
\partial_y^2 \psi\right) + \textrm{c.c.}\right] \right.  \nonumber \\
&\qquad \left. + \frac{2i e^2 \lambda_s}{N \eta \eta'} \dot{\phi} \left[ i \partial_y \psi^\dagger \psi + \textrm{c.c.}  \right]
+ \frac{is \lambda_s}{\eta^2} \phi\, \partial_x \left[i \partial_y\psi^\dagger \psi + \textrm{c.c.} \right] \right) \,.
\end{align}
This exercise shows explicitly how interactions cause currents to relax while conserving momentum. The coupling constants are order one at the fixed point.

\section{Specific heat of generalized phonons}
\label{sec:c}

The contribution of a `generalized phonon' with retarded Green's function $D^R(\omega,k)$ to the specific heat may be computed from the following general expression for the entropy density
\bea\label{eq:s}
s & = & \int \frac{d^dk}{(2\pi)^d} \frac{d\Omega}{\pi} \frac{\Omega}{8 T^2} \frac{\text{Im}\, \log D^R(\Omega,k)}{\sinh^2 \frac{\Omega}{2T}} \,, \\
& = & \int \frac{d^dk}{(2\pi)^d} \int_0^{\infty} \frac{d\Omega}{\pi} \frac{\Omega}{4 T^2} \frac{\arg D^R(\Omega,k)}{\sinh^2 \frac{\Omega}{2T}} \,. \label{eq:s2}
\eea
This formula is obtained from standard thermal field theory manipulations, using a spectral representation for $\log D^R(\omega,k)$. The specific heat is then given as usual by $c = T \pa s/\pa T$. To obtain the second line we used the fact that $D^R(-\Omega,k) = {\overline{D^R(\Omega,k)}}$.

Now consider a generalized phonon with spectral weight satisfying
\be\label{eq:genp}
\text{Im} \, D_\text{g-phon.}^R(\Omega,k) = 0 \qquad \text{for} \qquad |\Omega| > \Omega_\star(k) \,.
\ee
This is the simplest way to implement a Debye scale $\Omega_D = \max_k \Omega_\star(k)$. We are assuming nothing about the distribution of spectral weight at energies below $\Omega_\star(k)$. From the Krammers-Kronig relation we obtain the following result for the real part
\be
\text{Re} \, D_\text{g-phon.}^R(\Omega,k) < 0 \qquad \text{for} \qquad \Omega > \Omega_\star(k) \,.
\ee
This result uses only positivity of the spectral weight (for $\Omega > 0$). It follows that for $\Omega > \Omega_\star(k)$ the argument in (\ref{eq:s2}) satisfies $\arg D^R(\Omega,k) = \pi$. In the regime of temperatures where we obtain a linear in temperature scattering rate, $T \gg \Omega_D$, we can now isolate the following universal contribution to the entropy density
\be
s_\text{univ.} = \int \frac{d^dk}{(2\pi)^d} \int_{\Omega_\star(k)}^{T} \frac{d\Omega}{\Omega} = \int \frac{d^dk}{(2\pi)^d} \log \frac{T}{\Omega_\star(k)} \,.
\ee
This gives the constant specific heat
\be\label{eq:cgp}
c = \int \frac{d^dk}{(2\pi)^d} \,,
\ee
as we might have anticipated from energy equipartition.

\end{document}